\documentclass{PoS}

\usepackage{booktabs, threeparttable, stackengine}
\usepackage{array}
\newcolumntype{L}[1]{>{\raggedright\arraybackslash}p{#1}}
\setstackEOL{\#}
\setstackgap{L}{12pt}
\usepackage{array}
\newcolumntype{L}{>{\centering\arraybackslash}m{3cm}}

\emergencystretch=\maxdimen
\usepackage{natbib}
\usepackage{amsmath}        
\usepackage{amsfonts}       
\usepackage{amssymb}
\usepackage{color,graphicx}

\title{Fast Radio Bursts - implications and future prospects for Fermi}

\ShortTitle{FRBs - Implications and future prospects for Fermi}

\author{\speaker{Manisha Caleb}\\
        Jodrell Bank Centre for Astrophysics, School of Physics and Astronomy, University of Manchester, Manchester, M13 9PL, UK\\
        E-mail: \email{manisha.caleb@manchester.ac.uk}}

\abstract{The recent development of sensitive, high time
resolution instruments at radio telescopes has enabled the discovery of millisecond duration fast radio bursts
(FRBs). The FRB class encompasses a number of single pulses, many unique in their own way, so far hindering the development of a consensus for their origin. 
The key to demystifying FRBs lies in discovering many of them in realtime in order to localise them and identity commonalities.
Despite rigorous follow-up, only one FRB has been seen to repeat suggesting the possibility of there being two independent classes of FRBs and thus two classes of possible progenitors.  
This paper discusses recent developments in the field, the FRB-GRB connection, some of the open questions in FRB astronomy and how the next generation telescopes are vital in 
the quest to understand this enigmatic population. }

\FullConference{7th Fermi Symposium 2017\\
		15-20 October 2017\\
		Garmisch-Partenkirchen, Germany}

\begin{document}

\section{Introduction}

Time domain radio astronomy has recently discovered a new class of radio transient called Fast Radio Bursts (FRBs). 
This class comprises bright, millisecond duration ($\sim$ Jy) single pulses, mostly of singular occurrence,
exhibiting a quadratic frequency-dependent time delay (called dispersion measure), consistent with propagation through a cold ionized plasma.
The dispersion measure (DM) is the integrated free electron density along the line-of-sight given by,

\begin{equation}
\label{eq:dm}
\mathrm{DM} = \int_{0}^{d}{n_{e}dl},
\end{equation}

\noindent and can be approximated to be a proxy for distance. The DMs of pulsars in our Galaxy have been used to build a
model of the electron density distributions through various lines-of-sight \citep{Cordes, Yao}. 

\begin{figure}
\centering
\includegraphics[width=4.0 in]{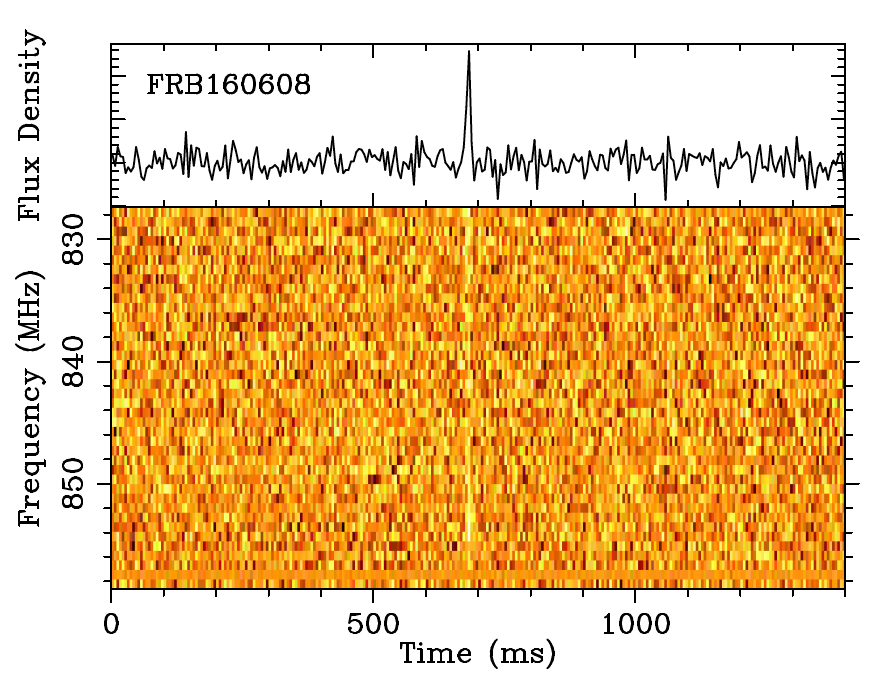}
\caption{Time vs frequency behaviour of FRB 160410 detected at the UTMOST. The top panel shows the time averaged profile and the bottom plot shows the frequency profile after the
effect of dispersion has been accounted for resulting in a width of 7.4 ms. Figure taken from \cite{Caleb}.}
\label{fig:dmwidth}
\end{figure}

The discovery of the sporadically emitting rotating radio
transients in 2006 \citep{McLaughlin} motivated astronomers to search for more such similar pulses resulting in the discovery of the prototypical FRB dubbed the
`Lorimer burst', in 2007 \citep{Lorimer}. The standout feature of the Lorimer burst was its 
dispersion measure, which was well in excess of the Galactic contribution for the observed line-of-sight. This is also the primary distinguishing feature between rotating radio transients and FRBs \citep{Keane}.
The redshift inferred from the excess DM, using a simple empirical scaling relation in \citet{Ioka} and \citet{Inoue} from gamma-ray bursts (GRBs),
implied an extragalactic if not cosmological ($\sim$ Gpc) origin for the Lorimer burst.
The inferred redshift is an upper limit, as the interstellar medium (ISM) of the host galaxy and environment local to the progenitor would also contribute to the DM. 
The high brightness temperatures of FRBs ($\mathrm{T_b} >10^{35}$ K) strongly
suggest a coherent emission mechanism \citep{Luan}. If FRBs are indeed coherent emitters at cosmological distances, they could open up
a whole new window to probe the extragalactic and even distant Universe. They could potentially be used to solve the case of the `missing baryons' in the Universe \citep{McQuinn}, map the intergalactic
magnetic field \citep{Zheng} and even acquire an independent measure of the dark energy equation of state \citep{Zhou}.

More than 30 FRBs have been discovered since 2007, with various telescopes (Parkes, Green Bank Telescope, Arecibo, UTMOST, ASKAP) over a range of frequencies 
(1.4 GHz, 800 MHz, 2 GHz, 4-8 GHz), of which only FRBs 24 have been published \citep{Lorimer, Thornton, Sarah, Ravi, Petroff, Spitler, Masui, Champion, NatKeane, RaviSci, Caleb, Petroff2, Bannister, Farah, Wael}. 
However only one of these 24 FRBs, the one discovered at the Arecibo observatory, has been seen to repeat, despite telescopes having spent several hundreds of hours re-observing the positions 
of known FRBs \citep{Rane}.
Leading progenitor models for FRBs range from binary neutron star mergers \citep{Totani, Kashiyama} and the collapse of a supramassive neutron star into a black hole \citep{Falcke} 
in the cataclysmic scenario, to giant flares from young energetic magnetars in
supernova remnants  \citep{Lyubarsky, Pen} and extragalactic pulsars in nearby galaxies \citep{Wasserman} in the repeating scenario.
Given the light travel-time and the fact that FRBs typically last only a few milliseconds, the radius of the source producing it would be $\sim$ few hundred kms implying a small emission region size.
This would not be the case though, if for example, FRBs were to arise at the intersection of a supernova remnant shock wave and progenitor source wind bubble \citep{Metzger}.

\section{Open questions in FRB astronomy}
It has been a decade since the discovery of FRBs and despite the rapid and significant development in the field, no consensus has emerged regarding their origin. Despite all the single pulses being broadly classified
as FRBs, no two pulses are the same.  In order to better understand the population, it is evident that we need a larger sample with arcsecond localisation and preferably
a counterpart at another wavelength. 

\subsection{Are there two (or more) classes of FRBs?}
The repeating FRB discovered at the Arecibo observatory is indicative of a type of progenitor that is not destroyed by the energetic events producing these FRBs. The repeat pulses have been
used to localise the source to a low-metallicity dwarf galaxy at $z \sim 0.19273(8)$ \citep{Chatterjee, Tendulkar}. This is the only FRB to have been unambiguously localised to a host.
The pulses appear to be highly clustered in time over the course of a couple of years with no underlying periodicity, and DMs consistent to within the uncertainties \citep{NatSpitler}. 
This is strongly suggestive of a single astronomical object being responsible for these bursts. No two pulses from the repeater look the same and they exhibit spectral cut-offs, frequency structure and wildly varying
spectral indices similar to the giant pulses from the Crab pulsar. 
An interesting property of these pulses is their lack of obvious pulse broadening due to possible multi-path scattering upon interaction with turbulent plasma along the path of propagation, quite commonly seen in pulsars. 
This suggests that the observed widths of $\sim 3 - 9$ ms are quite possibly their intrinsic widths \citep{Scholz}.

\begin{figure}
\centering
\includegraphics[width=4.1 in]{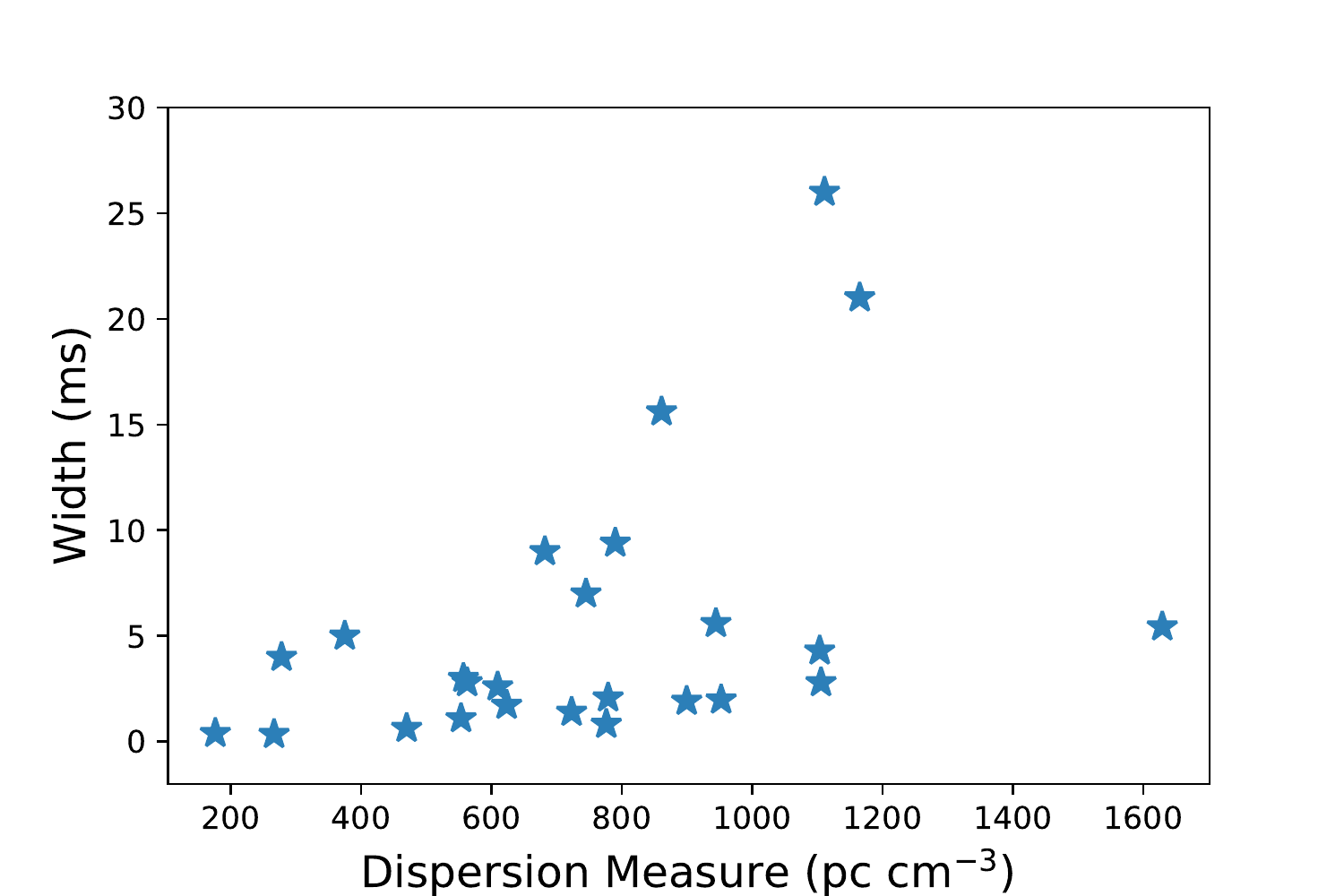}
\caption{The relationship between pulse width and DM. For the FRBs published, the pulse width is thought to be the result of Galactic (which is typically very small) and non-galactic (IGM $+$ host galaxy $+$ progenitor) contributions. 
If the scattering originates in the homogenous and diffuse IGM through which all FRBs propagate, all the pulses would be broadened, which is not the case. This suggests that the IGM is not responsible for the observed width making the host galaxy ISM and
progenitor environment strong candidates.}
\label{fig:dmwidth}
\end{figure}

The fact that none of the other FRBs have been seen to repeat yet, does not rule out the possibility of a non-cataclysmic progenitor. Most of the non-repeating FRBs have been discovered at the Parkes radio 
telescope whose sensitivity is $\sim 10$ times less than the Arecibo telescope. It should be noted that some of these pulses show spectral cut-offs and frequency structures similar to the repeater 
\cite[e.g.:][]{Champion, Wael}. The simultaneous detection 
of FRB 150418 with the Parkes radio telescope at 1.4 GHz and non-detection of the burst with the Murchison Widefield Array at 150 MHz, placed a limit of $\alpha \gtrsim -3.0$, on its spectral index \citep{NatKeane}.
Certain non-repeating FRB pulses do show strong evidence of multi-path scattering which when accounted for, imply much narrower intrinsic widths ($< 0.4 - 4$ ms) when compared to the repeater.

\begin{table}
\centering
\caption{Summary of the differences between the repeating and non-repeating FRBs.}
\label{tab:comparison}
\vspace*{5mm}
\centering
\begin{tabular}{|p{3.2cm}|p{5.4cm}|p{5.4cm}|} \hline
\textbf{Parameter} &        \textbf{Repeater (FRB 121102)} &   \textbf{Non-repeating FRBs}  \\ [1ex]  
\hline
Discovery telescope        & Arecibo   & Parkes, GBT, UTMOST, ASKAP      \\ 
\hline
Detection telescope  & GBT, VLA, Effelsberg & -- \\
\hline
Frequencies (GHz)	& 1.4, 2, $4-8$	& 1.4, 0.8, 0.843\\  
\hline
Spectral index	& $-10$ to $+15$	&  $\gtrsim -3.0$ for FRB150418 \\ 
\hline
Localisation	& Dwarf galaxy at $z \sim 0.19273(8)$	& No localisation yet \\ 
\hline
Polarisation	& No detectable polarisation	& Varied polarisation with no trend \\ 
\hline 
Widths 	& $3-9$ ms	& $\lesssim 0.4 - 26$ ms\\  
\hline
Scattering	         & No	& Measured for some FRBs\\ 
\hline
Frequency structure	& Yes	& Visible in some FRBs \\
\hline
Periodicity	           & No underlying periodicity 	& Singular events \\ 
\hline
DM variation	& Consistent to within uncertainties	& -- \\  
[1ex] 
\hline 
\end{tabular}
\end{table}

Figure \ref{fig:dmwidth} shows the relationship between pulse width and DM. The width of an FRB is the sum of contributions of Galactic and non-Galactic components. Pulsars at Galactic latitudes
similar to FRBs exhibit orders of magnitude smaller scattering timescales than FRBs \citep{Bhat, Krishnakumar}. 
The non-Galactic contributions could arise from the host galaxy and the intergalactic medium (IGM). \cite{Macquart} in their empirical scaling relation between DM and scattering, show that the IGM's contribution to the pulse 
broadening is orders of magnitude less than the Milky Way's ISM.
The FRB with the
greatest DM would be expected to have the broadest pulse width, which is evidently not the case from Figure \ref{fig:dmwidth}. 
Along with the fact that both time resolved and time unresolved pulses exist, this suggests that the IGM through which all the FRBs traverse, is not responsible 
for the pulse broadening, which makes the host galaxy and the progenitor circumburst medium strong candidates. 

In the simplest case, the repeater could belong to a different evolutionary phase of a given source population with a different log$N$-log$F$ relation compared to the non-repeating FRBs. 
The sensitivity of the Parkes radio telescope 
could constrain us from detecting the very faint pulses, resulting in only the bright tail end of the pulse energy distribution being visible, leading to the detection of one-off events. This is supported by the 
fact that even the published brightest pulse from the repeater would just be detectable above the threshold at Parkes. 
Or FRBs could simply have
multiple sub-populations, similar to GRBs.

\subsection{Emission in other wavebands? What are the prospects for Fermi?}
Of the 24 published FRBs, only the real-time detections have allowed for prompt follow-up at other wavelengths. Whether FRBs produce prompt signatures at other wavelengths still remains a mystery. 
Creating opportunities for follow-up and simultaneous observations with existing radio telescopes, high-energy and gravitational wave observatories in anticipation of the large sample of 
FRBs expected with next generation radio telescope could prove vital to progress in the field. We can expect breakthroughs in our
comprehension of FRBs by this approach, as was demonstrated in the case of GRBs and the repeating FRB 121102. The association of this FRB with a dwarf galaxy will potentially influence future follow-up
strategies. An association of an FRB with a counterpart at another wavelength would not only provide valuable insights regarding their progenitors but also their emission mechanisms. However this
type of multi-wavelength detection of FRBs at different wavelengths is highly uncertain as only one FRB has been seen to repeat. Commensal observing and shadowing similar to the strategy followed by the SUPERB
collaboration \citep{Superb}, offer great potential for the identification of an FRB progenitor. Most progenitor models for repeating FRBs do not predict emission at high energies as the progenitors are pulsar-like whose
emission is typically restricted to the radio spectrum \citep{Cordes2016}. Cataclysmic progenitors however do predict high energy emission in the form of gamma-rays, X-rays and optical afterglows \citep{Totani}.

It was recently reported that a gamma-ray transient detected with the burst alert telescope instrument on the SWIFT observatory was coincident with FRB 131102 
with an association significance of $3.2 \sigma$ \citep{DeLaunay}. 
The association claim was however refuted by \citet{Shannon} based on the non-detection of a radio afterglow at the location of the gamma-ray transient and the discovery of an AGN 
both temporally and spatially coincident with the FRB. Data taken with the Fermi gamma-ray burst monitor was searched for gamma-ray bursts associated with the repeating FRB 121102 during the time it 
was visible to the Fermi sky \citep{Scholz}. For the 4 radio bursts that were analysed, the corresponding gamma-ray data was found to be consistent with the persistent GBM background level. Based on the measured 
luminosity distance this corresponds to a 10-100 keV burst energy limit of $5 \times 10^{47}$ erg.

A targeted search for prompt radio emission from GRBs at 1.4 GHz was undertaken by \cite{BannisterGRB} using a 12-m telescope to slew automatically to the GRB coordinates based on GCN alerts. They report the
detection of 2 radio pulses, 524 and 1076 seconds after the GRB, with S/N $> 6\sigma$ and DMs greater than the Galactic contribution. However they calculate only a probability of 
2\% association with the corresponding GRBs
based on simple population arguments.
\citet{Palaniswamy} performed an experiment similar to \cite{BannisterGRB}. They observed five GRBs using a 26-m radio telescope, automated to respond to GCN alerts and slew to the source 
coordinates within minutes. Non-detections of any significant radio pulses $>5\sigma$ in their experiment, agree with the lack of consistency between the FRB and GRB event rates ($R_\mathrm{FRB} \sim 10^{-3} \, \mathrm{gal^{-1} \, yr^{-1}} > R_\mathrm{GRB} \sim 10^{-6} \mathrm{gal^{-1} \, yr^{-1}}$) presented in \citet{Thornton}.

\subsection{Can we probe cosmic magnetism?}
Faraday rotation of polarised sources have proven to be powerful probes of the magnetic field; both at the source and in the ISM. 
As the radio wave propagates from source to observer, the plane of polarisation of linearly polarised light is rotated
under the influence of a magnetic field with the magnitude of rotation quantified by the rotation measure (RM). The RM due to a source at cosmological distances is given by,

\begin{equation}
\mathrm{RM}(z) = \frac{e^3}{2\pi m_\mathrm{e}^{2} c^4} \int_0^z \! \frac{n_{e}(z) \, B_{||}(z)} {(1+z)^2} \cdot \frac{dl}{dz} \, dz \, \,\,\,\,\,\,\, \mathrm{rad \,m^{2}}
\label{eq:rm}
\end{equation}

\noindent where $n_\mathrm{e}$ is the electron density in particles per cubic metre, $B_{||}$ is the vector magnetic field parallel to the line-of-sight in microgauss, $dl$ is the elemental
vector towards the observer along the line-of-sight and $(1+z)$ is due to the redshifting of the observed frequencies. A combination of the RM and DM can be used to measure the integrated magnetic field along
the line-of-sight.
If FRBs are cosmological in nature, their polarisation properties could provide the first measurements of the intergalactic magnetic field and also insights into the magnetic fields associated with the progenitor.
Unlike the published pulses from the repeating FRB which show no polarisation, some non-repeating FRBs do show significant polarisation, though with considerable variation. 
Of the 24 FRBs, 8 have full polarisation information 
of which 6 are published. Though FRBs 140514 \citep{Petroff} and 150807 \citep{RaviSci} discovered at the Parkes radio telescope are highly linearly polarised, their RM values are consistent 
with Milky Way's foreground contribution for the
given lines-of-sight implying a negligible or zero contribution from the host galaxy and progenitor. In the host galaxy and progenitor, this is indicative of either low ordered magnetic fields or disordered 
magnetic fields which cancel each other resulting in no net RM value. The contribution from the IGM is typically of the order nano-Gauss \citep{Pshirkov} and negligible compared to the contributions from the Galactic foreground and
potential host galaxy and progenitor. In contrast are FRBs 110523 \citep{Masui} and 160102 (Caleb et al., in prep) which have significant linear polarisation and
RM values well in excess of the Galactic foreground contribution suggesting a highly magnetised medium and ordered magnetic fields in the host galaxy or immediate vicinity of the progenitor. 
The small sample size and lack of trend make it difficult to provide any explanation for FRB emission mechanisms and properties to date. 


\section{Summary and future prospects}
FRBs are one of the most tantalizing topics in astronomy of the last decade. We presently know of 38 FRBs of which only one is seen to repeat, but with no discernible periodicity. The large sample of non-repeating FRBs
is quite diverse with pulses that exhibit frequency structure, spectral cut-offs, are time resolved, time unresolved, linearly polarised, circularly polarised and unpolarised.  Multi-element interferometric detections of FRBs in the radio with near instantaneous localisation is the future. We can expect localisation to a few arcseconds radius with the next generation radio telescopes
like MeerKAT, UTMOST, Westerbork radio telescope, Australian Square Kilometre Array Pathfinder (ASKAP), Canadian Hydrogen Intensity Mapping Experiment (CHIME) and DSA-10, coming online. 
The UTMOST-2D project is a planned upgrade to the UTMOST telescope to provide better spatial localisation of a few arcseconds radius. The ASKAP and Westerbork telescopes with their
phased array feed technologies make them excellent instruments for FRB host localisations. The CHIME cylindrical telescope with its massive field-of-view of 250 deg$^{2}$ is expected to detect 
tens of FRBs per day though
localisation will be relatively poor ($\sim$ arcmin). The MeerKAT pathfinder to the SKA is also expected to provide near instantaneous localisation of a few arcseconds radius. The MeerTRAP project at MeerKAT 
will undertake high time resolution, fully commensal transient searches in parallel with most of the legacy science project observations. 
This is particularly advantageous for FRB searches as it will allow a large sky coverage, multiple visits to the same field to search for repeats, and variable cadence observations.

These future radio telescopes will enable unambiguous association of an FRB with a host galaxy even in the case of non-repeating bursts. 
Rapid follow-up and high cadence observations at X-ray, optical and gamma-ray energies accompanying the localisation in the radio is required to robustly sample the lightcurve of a transient afterglow.
The current limits are insufficiently stringent to constrain the FRB-GRB connection. However future radio telescopes along with sensitive observations with Fermi and the Cerenkov Telescope Array
might be able to constrain the connection. We can expect the next decade to answer some if not all the currently open questions in FRB
astronomy.

\section{Acknowledgements}
This project has received funding from the European Research Council (ERC) under the European Union's Horizon 2020 research and innovation programme (grant agreement No 694745).
Parts of this research was funded by the Australian Research Council Centre for All-Sky Astrophysics (CAASTRO), through project number CE110001020, and the Laureate Fellowship FL150100148.

\end{document}